\documentclass[preprint,preprintnumbers,superscriptaddress]{revtex4}
\usepackage{graphicx}
\usepackage{dcolumn}
\usepackage{bm}
\usepackage{color}
\usepackage{amsfonts}
\usepackage{mathrsfs}
\usepackage{amsmath,amssymb}

 \usepackage[colorlinks,linkcolor=blue,anchorcolor=blue,citecolor=blue]{hyperref}

\newcommand{\beq}{\begin{equation}}
\newcommand{\eeq}{\end{equation}}
\newcommand{\beqn}{\begin{eqnarray}}
\newcommand{\eeqn}{\end{eqnarray}}
\newcommand{\bsub}{\begin{subequations}}
\newcommand{\esub}{\end{subequations}}

\begin{document}

\title{Systematic study of symmetry energy coefficient in finite nuclei}

\author{H. Mei}
\address{School of Physical Science and Technology, Southwest
University, Chongqing 400715, China}

\author{Y. Huang}
\address{School of Physical Science and Technology, Southwest
University, Chongqing 400715, China}
\address{School of Physics, Peking University,
Beijing 100871, China}

\author{J. M. Yao}
\address{School of Physical Science and Technology, Southwest
University, Chongqing 400715, China}

\author{H. Chen}
\address{School of Physical Science and Technology, Southwest
University, Chongqing 400715, China}

\begin{abstract}

The symmetry energy coefficients in finite nuclei have been studied
systematically with a covariant density functional theory (DFT) and
compared with the values calculated using several available mass
tables. Due to the contamination of shell effect, the nuclear
symmetry energy coefficients extracted from the binding energies
have large fluctuations around the nuclei with double magic numbers.
The size of this contamination is shown to be smaller for the nuclei
with larger isospin value. After subtracting the shell effect with
the Strutinsky method, the obtained nuclear symmetry energy
coefficients with different isospin values are shown to decrease
smoothly with the mass number $A$ and are subsequently fitted to the
relation $\dfrac{4a_{\rm
sym}}{A}=\dfrac{b_v}{A}-\dfrac{b_s}{A^{4/3}}$. The resultant volume
$b_v$ and surface $b_s$ coefficients from axially deformed covariant
DFT calculations are $121.73$ and $197.98$ MeV respectively. The
ratio $b_s/b_v=1.63$ is in good agreement with the value derived
from the previous calculations with the non-relativistic Skyrme
energy functionals. The coefficients $b_v$ and $b_s$ corresponding
to several available mass tables are also extracted. It is shown
that there is a strong linear correlation between the volume $b_v$
and surface $b_s$ coefficients and the ratios $b_s/b_v$ are in
between $1.6-2.0$ for all the cases.

\end{abstract}


\maketitle

\section{Introduction}
\label{intro}

 The advent of radioactive ion beams (RIBs)~\cite{Tanihata85PRL,Bertulani01}
 provides a useful tool for studying exotic nuclei far from the $\beta$-stability.
 Hitherto, RIBs have already disclosed many structure phenomena in
 nuclei with extreme isospin values, and the next generation of radioactive-beam
 facilities are expected to produce more and more exotic
 nuclei~\cite{Mueller93,Tanihata95,Hansen95,Casten00,Mueller01,Jonson04,Jensen04}.
 These exotic nuclei provide us useful information on the equation of
 state (EOS) of asymmetric nuclear matter, which has not been well determined but is important
 for understanding both the structure of unstable nuclei and the
 properties of neutron stars. One of the most important quantities
 is the nuclear symmetry energy, which affects significantly the binding
 energy and radii of neutron-rich nuclei~\cite{Sumiyoshi93}.
 Furthermore, the chemical composition,
 the evolution of lepton profiles, cooling process and the neutrino fluxes in neutron stars
 depend strongly on the nuclear symmetry energy~\cite{Sumiyoshi95,Chasman03}.

 Many investigations have already been carried out to study the
 density and isospin dependence of symmetry energy for nuclear matter~\cite{Danielewicz03,Baran05,Li08}.
 The EOS for asymmetric nuclear matter has a general parabolic form
 $\dfrac{E}{A}(\rho,\beta)=\dfrac{E}{A}(\rho)+E_{\rm sym}(\rho, 0)\beta^2+O(\beta^2)$
 with baryon density $\rho=\rho_n+\rho_p$, isospin asymmetry $\beta=(\rho_n-\rho_p)/\rho$,
 and nuclear symmetry energy coefficient $E_{\rm sym}(\rho)=\dfrac{1}{2!}
 \dfrac{\partial^2 E(\rho,\beta)}{\partial \beta^2}\vert_{\beta=0}$.
 Near the nuclear saturation density, the $E_{\rm sym}(\rho)$ has a
 strong influence on the neutron density distribution and thus
 on the neutron skin in stable and exotic nuclei. Previous studies have demonstrated that
 the neutron skin size can yield information on the derivative of symmetry
 energy with respect to density~\cite{Oyamatsu98,Brown00,Typel01,Furnstahl02,Centelles09,Roca-Maza11}.

 In recent years, the extraction of symmetry energy from finite nuclei has
 attracted much attention. According to the semi-empirical mass formula of liquid drop model (LDM),
 the symmetry energy in finite nuclei is proportional to the square of the difference between neutron and
 proton numbers. However, recent studies indicate that the symmetry energy is more reasonable to be
 parameterized as $E_{\rm sym} \sim T (T + c)$
 with $T=\vert N-Z\vert/2$~\cite{Vogel00}. The empirical fits
 in Ref.~\cite{Janecke65} show a clear preference for $c=1$.
 The fits with either $c=0$ as in the LDM, or with $c=4$ as in the Wigner SU(4) symmetry are disfavored.
 Similar conclusion has also been reached in Ref.~\cite{Jensen84}.
 Later on, the symmetry energy of finint nuclei has been examined with several microscopic models, including
 the Hartree-Bogoliubov approach with random phase approximation~\cite{Neergard02},
 and energy density functional (EDF) of Skyrme force~\cite{Satula03} or the relativistic
 meson-exchange interaction~\cite{Ban06}, all of which favor $c=1$ as well.

 On the other hand, it has been shown that the surface tension needs to depend on asymmetry.
 This dependence modifies the surface energy and implies the emergence of asymmetry skin~\cite{Danielewicz03}. It
 was pointed out that the surface symmetry term was required in the symmetry
 energy coefficient, namely, $\dfrac{4a_{\rm
sym}}{A}=\dfrac{b_v}{A}-\dfrac{b_s}{A^{4/3}}$, in order to describe
the energies of light asymmetric nuclei at the level similar to the
other nuclei. Its physical origin is traditionally explained in
terms of the kinetic energy and mean isovector potential
contributions respectively~\cite{Bohr69}. In the previous studies,
the ratio of the surface-to-volume contributions to the symmetry
energy coefficient $r_{S/V}(\equiv b_s/b_v)$ has been estimated from
the electric dipole strength distribution using the hydrodynamical
model~\cite{Lipparini82}, or from the experimental
masses~\cite{Moller95,Mukhopadhyay07,Kirson08,Wang10}, mass and
radii~\cite{Danielewicz03,Dieperink07}, or the excitation energies
of isobaric analog states~\cite{Danielewicz07}. Recently, an attempt
has also been made to extract symmetry energy and the ratio
$r_{S/V}$ from the separation energies through the displacement of
neutron and proton chemical potentials~\cite{Kolomietz10}.

The density functional theory (DFT) in nuclear physics is nowadays
the most important microscopic approach for large-scale nuclear
structure calculations in medium-heavy and heavy nuclei and it has
been successfully employed for the description of nuclei around and
far from the $\beta$-stability~\cite{Bender03}. In the framework of
DFT~\cite{Reinhard06}, the LDM parameters and effective symmetry
energy coefficient were extracted for nuclei with huge numbers of
nucleons, of the order of $10^6$. In the mean time, the systematical
extraction of symmetry energy coefficient from realistic finite
nuclei, for which, the surface energy term is important, was carried
out based on several non-relativistic Skyrme EDFs~\cite{Satula06},
where the global mass dependence of symmetry energy coefficient was
studied by switching off the Coulomb and pairing effects. It has
been found that the ratio of the surface-to-volume contributions to
the symmetry energy coefficient, $r_{S/V}(\equiv b_s/b_v)$ is around
$1.6$, which is consistent with the value of Ref.~\cite{Moller95},
but much different from the estimation $2.0\leq r_{S/V}\leq2.8$ in
Ref.~\cite{Danielewicz03}. However, the recent careful study with
the considerations of Coulomb interaction and shell effect in
Ref.~\cite{Nikolov11} shows that this ratio could be quite different
for different Skyrme forces, for instance, $r_{S/V} = 1.2$ for BSk6,
1.7 for SkM*, and 2.2 for SkI3.

In the past decades, nuclear covariant DFT has achieved comparable
success with the non-relativistic DFT in the description of ground
state properties of both spherical and deformed nuclei all over the
nuclear chart~\cite{Reinhard89,Ring96,Vretenar05,Meng06}. In
particular, there are many advantages using the EDF with manifest
covariance, including natural inclusion of the nucleon spin degree
of freedom, automatical emergence of nuclear spin-orbit potential,
an unique parametrization of time-odd components (nuclear currents
and magnetism) in nuclear mean-field as well as the natural
saturation mechanism for nuclear matter.

In recent several years, the covariant DFT theory with
point-coupling interaction has attracted much more attention. It
shows great advantages in the extension for nuclear low-lying
excited states by using projection techniques~\cite{Yao09},
generator coordinate methods~\cite{Niksic06,Yao10,Yao11,Yao11-2} and
collective Hamiltonian~\cite{Niksic09}. In this framework, the
PC-PK1 set~\cite{Zhao10} was proposed most recently by fitting to
the binding energies of 60 selected spherical nuclei and the charge
radii of 17 selected spherical nuclei from $\rm O$ to $\rm Pb$
isotopes. The success of PC-PK1 has been illustrated in the
description of infinite nuclear matter and finite nuclei for both
ground-state and low-lying excited states. Furthermore, the PC-PK1
provides a good description for the isospin dependence of nuclear
binding energy along either isotopic or isotonic chain, which is
particular important for description of nuclear symmetry energy in
finite nuclei.  In view of these facts, in this paper, we will study
the symmetry energy coefficient of realistic finite nuclei
systematically using both spherical and axially deformed RMF
approaches with the PC-PK1 covariant EDF, where the shell correction
energy will be subtracted using the Strutinsky method~\cite{Ring80}.
The obtained symmetry energy coefficient will be compared with those
extracted from the binding energies of several available mass
tables. Compared with our previous work with the PK1 effective
interaction~\cite{Mei09}, in this work, we make a lot of
improvements, including the consideration of shell correction
energy, axial deformation, pairing correlation and Coulomb
interaction in the RMF calculations, which might have influences on
the extracted symmetry energy of realistic nuclei.

The paper is organized as follows. In Sec.~\ref{SecII} we briefly
describe the method used to extract the nuclear symmetry coefficient
in finite nuclei. The results and discussion will be given in
Sec.~\ref{SecIII}. Finally, a summary is presented in
Sec.~\ref{SecIV}.

 \section{The method}
 \label{SecII}
 \subsection{The relativistic mean-field approach with point-coupling interaction}

 The relativistic mean-field (RMF) approach with point-coupling
 interaction for nucleons has been described in detail in Refs.~\cite{Yao09,Zhao10}.
 Here, we present only the outline of this approach. It starts from the following Lagrangian density,
 \beqn
  \label{Lagrangian}
   {\cal L}
   &=& \bar\psi(i\gamma_\mu\partial^\mu-m)\psi\nonumber\\
             &&-\frac{1}{2}\alpha_S(\bar\psi\psi)(\bar\psi\psi)
                -\frac{1}{2}\alpha_{V}(\bar\psi\gamma_\mu\psi)(\bar\psi\gamma^\mu\psi)\nonumber\\
              &&  -\frac{1}{2}\alpha_{TV}(\bar\psi\vec\tau\gamma_\mu\psi)(\bar\psi\vec\tau\gamma^\mu\psi))\nonumber\\
             &&-\frac{1}{2}\delta_S\partial_\nu(\bar\psi\psi)\partial^\nu(\bar\psi\psi)
             -\frac{1}{2}\delta_V\partial_\nu(\bar\psi\gamma_\mu\psi)\partial^\nu(\bar\psi\gamma^\mu\psi)\nonumber\\
             &&
             -\frac{1}{2}\delta_{TV}\partial_\nu(\bar\psi\vec\tau\gamma_\mu\psi)\partial^\nu(\bar\psi\vec\tau\gamma^\mu\psi)\nonumber\\
             &&
              -\frac{1}{3}\beta_S(\bar\psi\psi)^3-\frac{1}{4}\gamma_S(\bar\psi\psi)^4
                   -\frac{1}{4}\gamma_V[(\bar\psi\gamma_\mu\psi)(\bar\psi\gamma^\mu\psi)]^2\nonumber\\
             &&
             -\frac{1}{4}F^{\mu\nu}F_{\mu\nu}-e\bar\psi\gamma^\mu\dfrac{1-\tau_3}{2}\psi A_\mu,
 \eeqn
 which contains 9 coupling constants
 $\alpha_S$, $\alpha_V$, $\alpha_{TV}$, $\beta_S$, $\gamma_S$, $\gamma_V$, $\delta_S$,
 $\delta_V$ and $\delta_{TV}$. The subscripts indicate the symmetry of the couplings:
 $S$ stands for scalar, $V$ for vector, and $T$ for isovector, while the symbol refer
 to the additional distinctions: $\alpha$ refers to four-fermion term, $\delta$ to
 derivative couplings, and $\beta$ and $\gamma$ to the third- and fourth-order
 terms, respectively.

 Using the mean-field approximation and the ``no-sea''
 approximation,  one finds the energy density functional corresponding to the Lagrangian density (\ref{Lagrangian}).
 Minimization of the energy density functional with respect to $\bar\psi_k$
 gives rise to the Dirac equation (i.e., Kohn-Sham equation) for the single nucleon
 wave function $\psi_k$,
 \beqn
  \label{DiracEq}
  [\gamma_\mu(i\partial^\mu-V^\mu)-(m+S)]\psi_k=0.
 \eeqn
 The single-particle effective Hamiltonian contains local scalar $S(\bm{r})$ and vector
 $V^\mu(\bm{r})$ potentials
 \beq
 \label{potential}
   S(\bm{r})    =\Sigma_S, \quad
   V^\mu(\bm{r})=\Sigma^\mu+\vec\tau\cdot\vec\Sigma^\mu_{TV},
 \eeq
 where the nucleon scalar-isoscalar $\Sigma_S$, vector-isoscalar $\Sigma^\mu$ and
 vector-isovector $\vec\Sigma^\mu_{TV}$ self-energies are given in terms of the various
 densities and currents,
 \bsub\beqn
  \Sigma_S           &=&\alpha_S\rho_S+\beta_S\rho^2_S+\gamma_S\rho^3_S+\delta_S\triangle\rho_S,\\
  \Sigma^\mu         &=&\alpha_Vj^\mu_V +\gamma_V (j^\mu_V)^3
                       +\delta_V\triangle j^\mu_V + e A^\mu,\\
  \vec\Sigma^\mu_{TV}&=& \alpha_{TV}\vec j^\mu_{TV}+\delta_{TV}\triangle\vec j^\mu_{TV}.
 \eeqn\esub
The local densities and currents are defined by,
\bsub%
\label{currents}
\beqn%
  \label{E13a}
  \rho_S(\bm{r})          &=&\sum_{k }v^2_k\bar\psi_k(\bm{r})\psi_k(\bm{r}),\\
  \label{E13b}
  j^\mu_{V}(\bm{r})       &=&\sum_{k }v^2_k\bar\psi_k(\bm{r})\gamma^\mu\psi_k(\bm{r}),\\
  \label{E13c}
  \vec j^\mu_{TV}(\bm{r}) &=&\sum_{k
  }v^2_k\bar\psi_k(\bm{r})\vec\tau\gamma^\mu\psi_k(\bm{r}),
\eeqn \esub
 where the summation $\sum\limits_{k}$ runs over only
 positive-energy states with the occupation probabilities $v^2_k$.

 For ground state of an even-even nucleus one has time reversal
 symmetry. The space-like components of the currents and
 the spatial part of the vector potential vanish. Moreover, because of charge conservation in nuclei,
 only the 3rd-component of isovector potential $\vec\Sigma^\mu_{TV}$ contributes.
 The Coulomb field $A_0$ is determined by Poisson's equation.

 Pairing correlations between nucleons are treated in the BCS approximation,
 where the density-independent $\delta$-force is used in the particle-particle channel.
 Therefore, the nuclear total energy is contributed from both particle-hole and particle-particle channels.
 Moreover, the center-of-mass (c.m.) correction to the total energy is taken into account microscopically,
 \beq
  \label{Eq:Ecm}
  E^{\rm mic}_{\rm cm}=-\dfrac{1}{2mA}\langle\hat P^{2}_{\rm cm}\rangle,
 \eeq
 where $m$ is the mass of nucleon, and $A$ is mass number.
 $\hat P_{\rm cm}=\sum_i^A \hat p_i$ is the total momentum in the
 c.m. frame.

\subsection{Extraction of symmetry energy coefficient}
 In the conventional semi-empirical mass formula of LDM for the binding
 energy of nuclei, the symmetry energy is given
 in terms of the isospin value $T$~\cite{Ring80},
 \begin{equation}
 \label{Esym}
 E_{\rm sym}(A,T)
 =\dfrac{4a_{\rm sym}}{A}(A,T) T^2,
 \end{equation}
 where $A$ is the mass number. $\dfrac{4a_{\rm sym}}{A}(A,T)$ represents the
 symmetry energy coefficient. The isospin value $T$ is given by $\vert T\vert=\vert N-Z\vert/2$.
 Recent microscopic studies~\cite{Duflo95,Satula03,Ban06,Neergard02,Janecke65} with
 the consideration of nuclear Wigner energy showed that the nuclear
 symmetry energy is more proper to relate the isospin value $T$ by the following relation,
 \begin{equation}
 \label{SEC}
 E_{\rm sym}(A,T) = \dfrac{4a_{\rm sym}}{A}(A,T) T(T + 1).
 \end{equation}
 On the other hand, according to the mass formula of LDM, the difference between binding energies
 of isobaric nuclei with the same odd-even parity is only related to the
 Coulomb energy and symmetry energy terms,
 \begin{eqnarray}
   B(A, T)-B(A, 0)= E_{\rm Coul.}(A, T)- E_{\rm Coul.}(A, 0) + E_{\rm sym}(A, T)- E_{\rm sym}(A, 0),
 \end{eqnarray}
 where $B(A, 0)$ is the total binding energy of $N=Z$ isobaric nucleus
 with the mass number $A$.
 Subsequently, the nuclear symmetry energy in Eq.(\ref{SEC}) is simply determined by subtracting the
 contribution of Coulomb energy from the nuclear total binding energy as,
 \begin{eqnarray}
 \label{Exp_NSE}
   E_{\rm sym}(A,T)=[B(A, T)-B(A, 0)]-[E_{\rm Coul.}(A, T) - E_{\rm Coul.}(A, 0)].
  \end{eqnarray}
 Combining Eqs.(\ref{SEC}) and (\ref{Exp_NSE}), one
 can determine the value of symmetry energy coefficient if one knows the
 total binding energy and Coulomb energy of the concerned nucleus.
 The shell correction energy can contaminate the symmetry energy
 obtained in this way and therefore should be subtracted from the binding energy.

 \section{Results and discussion}
 \label{SecIII}

In the RMF calculations, we adopt the recent parameterized
relativistic point-coupling interaction PC-PK1~\cite{Zhao10}, which
has been adjusted to the binding energies and charge radii of
spherical nuclei from $\rm O$ to $\rm Pb$ isotopes. With the
restriction of spherical/axial symmetry, the Dirac equation for
nucleons is solved by expanding the Dirac spinor $\psi_k$ on a set
of harmonic oscillator basis with $20/16$ shells. Pairing
correlations between nucleons are treated with the BCS approximation
using a density-independent $\delta$ force. The pairing strengths
have been adjusted to fit the average neutron pairing gaps in
$^{122}\rm Sn$, $^{124}\rm Sn$ $^{200}\rm Pb$ and the average proton
gaps in $^{92}\rm Mo$, $^{136}\rm Xe$, and $^{144}\rm Sm$.  More
details about the numerical calculations can be found in
Ref.~\cite{Zhao10}.

 \begin{figure}[]
 \includegraphics[width=12cm]{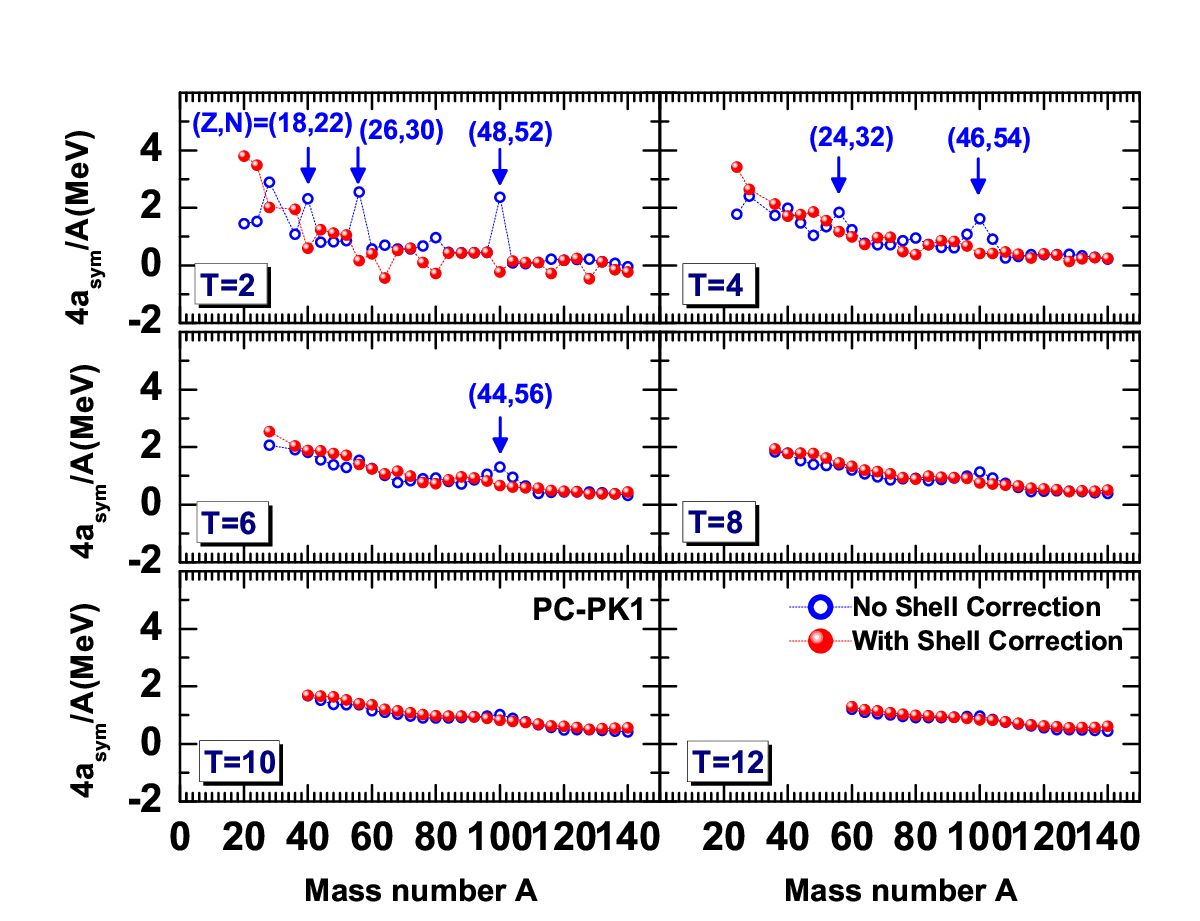}
 \caption{The nuclear symmetry energy coefficients for finite nuclei with $8\leq Z\leq70$ and $T=2,4,\cdots,12$
 in the spherical relativistic mean-field calculations with effective interaction PC-PK1 as functions of the
 corresponding mass number $A$. The filled (open) symbols
 represent the values with (without) taking into account the shell correction.}
 \label{fig1}
\end{figure}

 Figure~\ref{fig1} displays the nuclear symmetry energy coefficient
 for finite nuclei with $8\leq Z\leq70$ and $T=2, 4, \cdots, 12$
 from the spherical RMF calculations
 with effective interaction PC-PK1~\cite{Zhao10} as a function of the
 corresponding mass number $A$. The filled (open) symbols
 represent the values with (without) taking into account the shell correction energies that
 are calculated with the Strutinsky method~\cite{Ring80}. It shows clearly that
 due to the contamination of shell correction energy, the resultant nuclear symmetry energy
 coefficients have large fluctuations around the nuclei with double magic numbers.
 Moreover, the amplitude of these fluctuations decreases with
 the isospin value $T$. It indicates that the symmetry energy becomes less sensitive to the
 shell effect for the nuclei with larger isospin value $T$.
 After subtracting the shell effect energy evaluated with the Strutinsky method,
 the obtained nuclear symmetry energy coefficients are smoothly decreasing with the mass number $A$.

\begin{figure}[]
 \includegraphics[width=12cm]{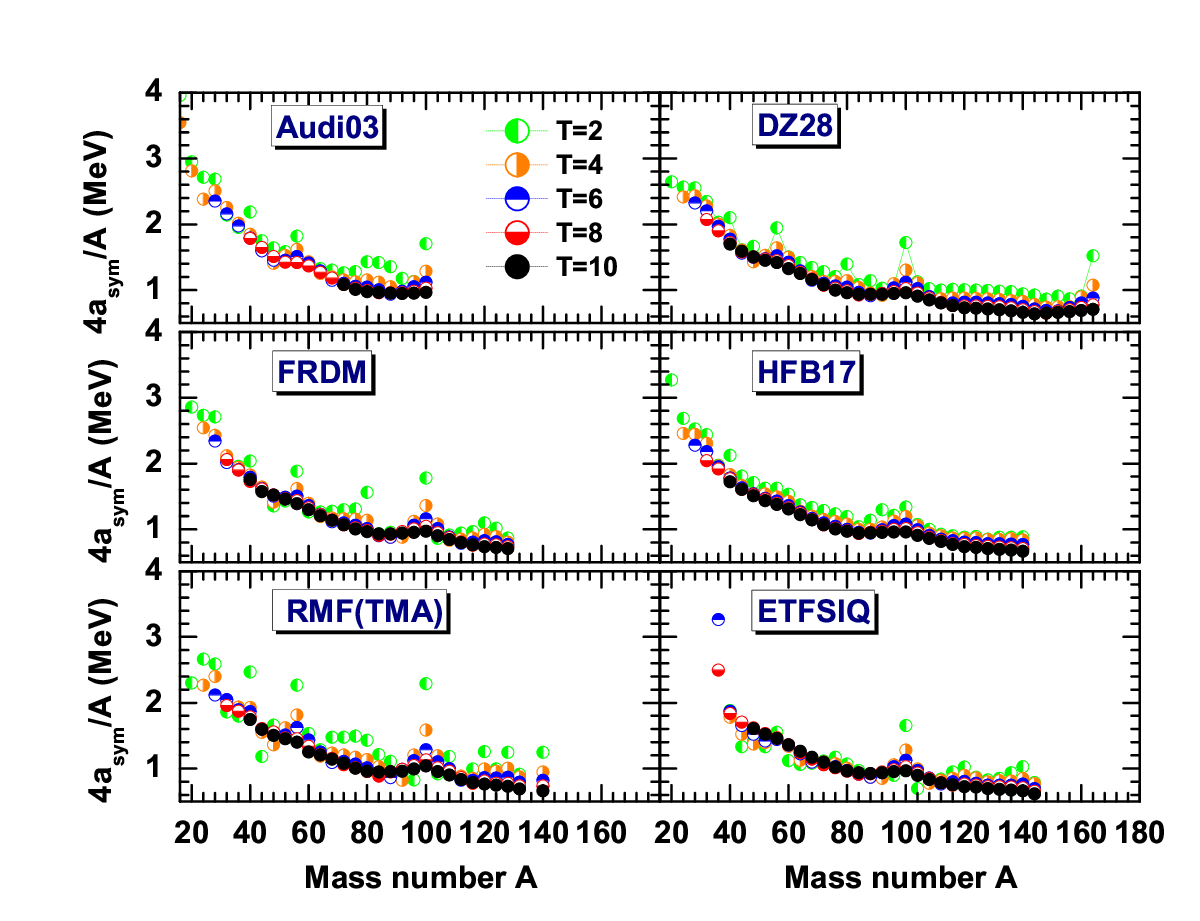}
 \caption{The nuclear symmetry energy coefficient for finite nuclei
 with $T=2, 4, \cdots, 10$ respectively in different mass tables as a function of the
 corresponding mass number $A$. }
 \label{fig2}
 \end{figure}

Furthermore, the nuclear symmetry energy coefficients are calculated
using the binding energy of nuclei with $T=2,4,\cdots, 10$ from
several available mass tables, including the ``DZ28"~\cite{Duflo95},
``FRDM"~\cite{Moller95}, ``HFB17"~\cite{Goriely09},
``RMF(TMA)"~\cite{Geng05}, ``ETFSIQ"~\cite{Pearson96} as well as the
experimental data ``Audi03"~\cite{Audi03}. In the calculations, we
take the Coulomb energy $E_{\rm Coul.}(A,T)$ of a specific nucleus
as the value of an uniformly charged droplet, i.e.,
  \begin{equation}
  \label{Eq2}
   E_{\rm Coul.}(A,T)
   =\dfrac{3}{5}\dfrac{Z^2e^2}{R_c}[1-\dfrac{5}{4}(\dfrac{3}{2\pi})^{2/3}
   \dfrac{1}{Z^{2/3}}],~~R_c=1.2A^{1/3}.
  \end{equation}
 Figure~\ref{fig2} shows the resultant symmetry energy coefficient from different mass tables as a function of the
 mass number $A$.  Since the shell effect is not subtracted from the binding energies of these mass tables,
 similar fluctuation as those in Fig.~\ref{fig1} (open symbols) is observed in Fig.~\ref{fig2}.

 \begin{figure}[]
 \includegraphics[width=12cm]{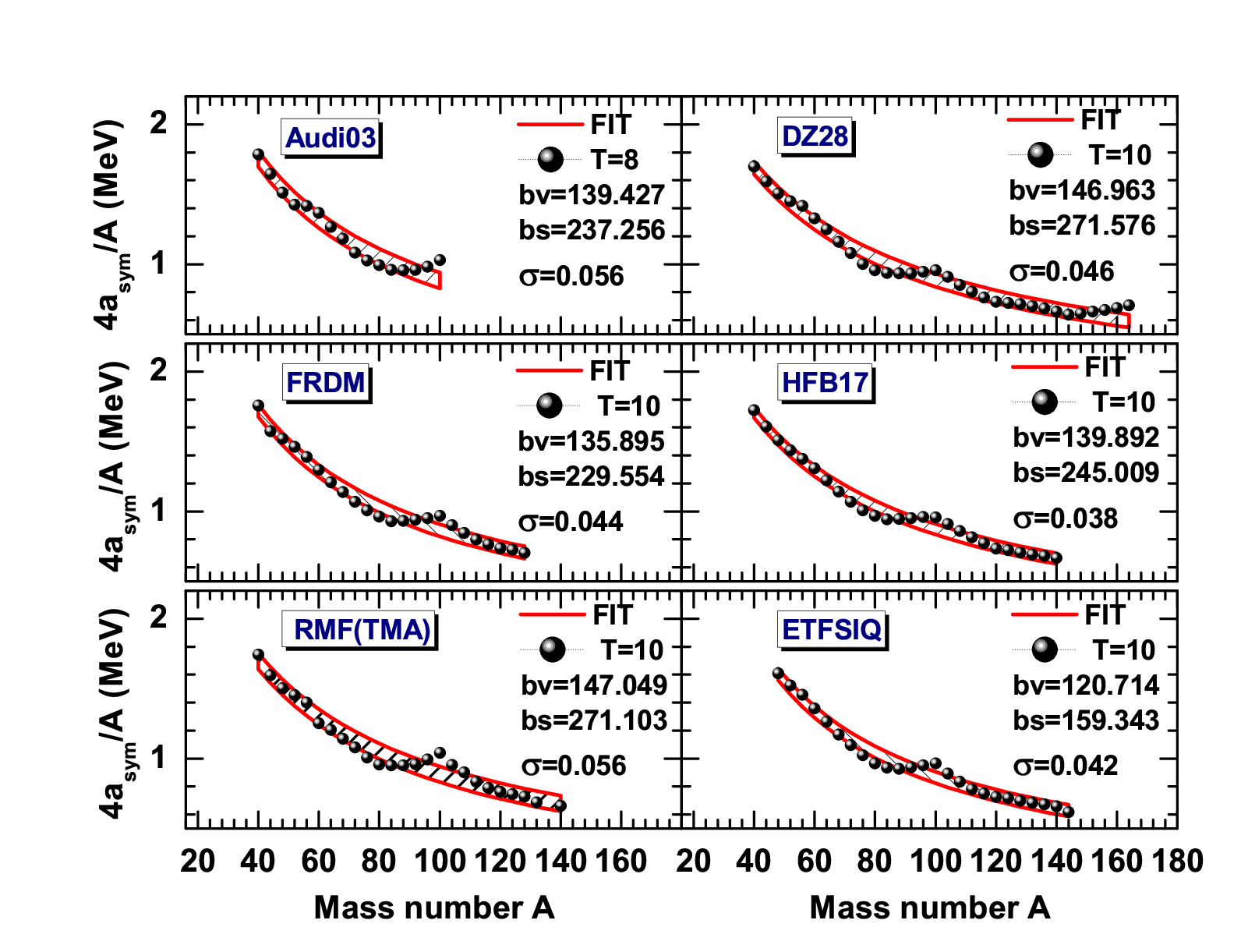}
 \caption{The nuclear symmetry energy coefficient extracted from
 finite nuclei with $T=10$ in different mass tables.
 The optimal values of volume coefficient $b_v$ and surface coefficient $b_s$ fitted to the relation (\ref{secp})
 and the corresponding standard root-mean-square errors $\sigma$ are presented.
 The shade area indicates the $\dfrac{4a_{\rm sym}}{A}\pm1\sigma$, where $\dfrac{4a_{\rm
 sym}}{A}$ is calculated using the corresponding optimal values of $b_v$ and $b_s$.}
 \label{fig3}
 \end{figure}

 \begin{figure}[]
 \includegraphics[width=10cm]{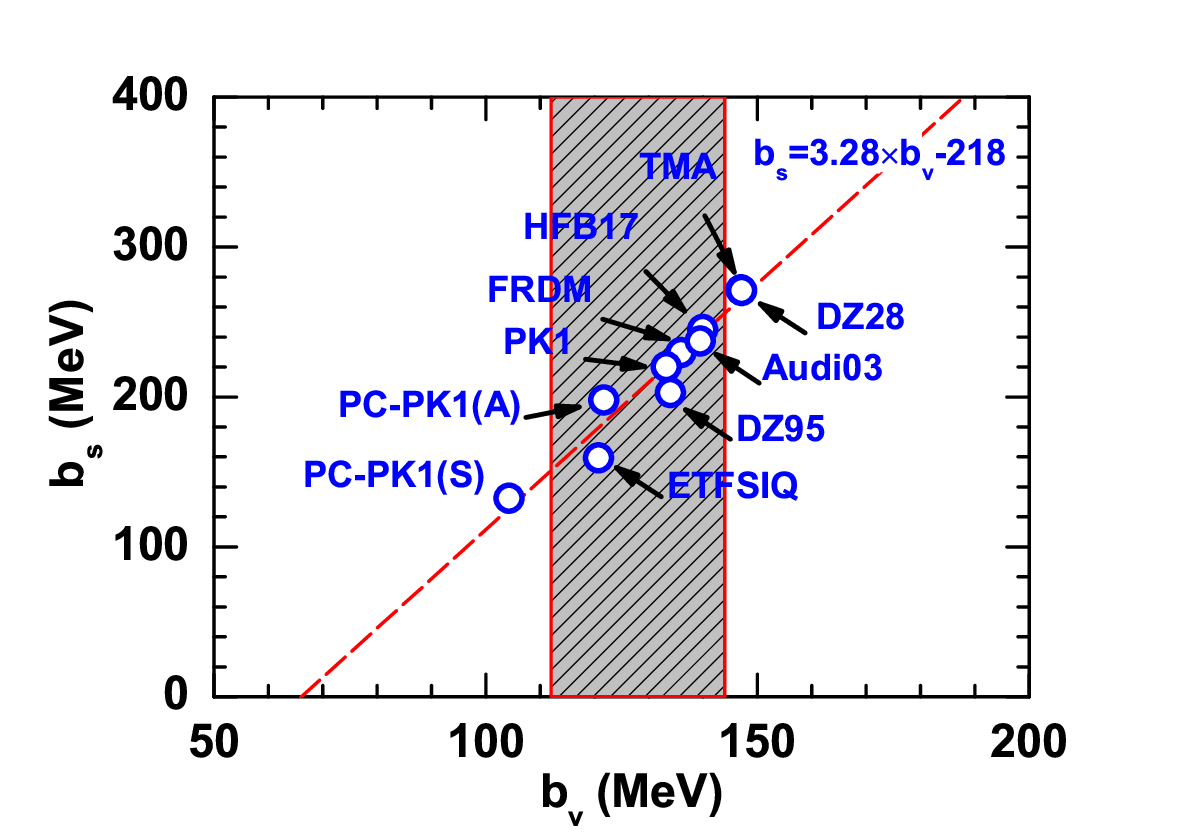}
 \caption{The surface coefficient $b_s$ as a function of the volume coefficient $b_v$.
 The shade area indicates the empirical value of $b_v=128\pm16$ (MeV) for infinite nuclear
 matter (see text for details).}
 \label{fig4}
 \end{figure}

 The symmetry energy coefficient of finite nucleus with mass number $A$ is usually parameterized as follows,
 \begin{eqnarray}
 \label{secp}
  \dfrac{4a_{\rm sym}}{A}=
       \dfrac{b_v}{A}-\dfrac{b_s}{A^{4/3}}.
 \end{eqnarray}
In Ref.~\cite{Duflo95}, it was found that $b_v=134.4$, $b_s=203.6$
MeV, while in Ref.~\cite{Ban06}, the axial RMF calculations with the
PK1~\cite{Long04} interaction for the $A = 40$, $48$, $56$, $88$,
$100$, $120$, $140$, $160$, $164$, and $180$ isobaric chains gave
$b_v=133.2$ and $b_s=220.3$ MeV. We note that in Ref.~\cite{Ban06},
the Coulomb interaction between protons was switched off and the
contamination of shell effect was not considered in the
calculations. Since the shell effect in nuclear symmetry energy
coefficient is much weaker for high isospin value as shown in
Fig.~\ref{fig1} and Fig.~\ref{fig2}, we perform a global fit of the
symmetry energy coefficients to the relation (\ref{secp}) via the
Levenberg-Marquardt method. Considering the limited experimental
data for nuclei with $T=10$, the corresponding experimental value
``Audi03" extracted from the finite nuclei with $T=8$ is also shown
in Fig.~\ref{fig3}. The obtained $b_v$ and $b_s$ values, together
with the standard root-mean-square errors $\sigma$ are presented.
The small $\sigma$ value indicates the good quality of the fits. The
resultant volume coefficient $b_v$ and surface coefficient $b_s$ are
plotted in Fig.~\ref{fig4}, where the shade area indicates the
empirical value of $b_v=128\pm16$ MeV (corresponding to $a_{\rm
sym}=32\pm4$ MeV for infinite nuclear matter). In addition, the
values of $b_v$ and $b_s$ from the axial RMF calculations with the
PK1~\cite{Long04} effective interactions (labeled as ``PK1") from
Ref.~\cite{Ban06}, as well as the spherical/axial RMF calculations
with the PC-PK1 effective interaction (labeled as
``PC-PK1(S)/PC-PK1(A)") from this work are plotted for comparison.
It is shown that all the obtained $b_v$ values are in good agreement
with the empirical value, except the case of spherical RMF
calculations with the PC-PK1 interaction, which gives $b_v=104.25$
and $b_s=132.38$ MeV. After taking into account the effect of axial
deformation, the obtained $b_v$ value of ``PC-PK1(A)" agrees well
with the empirical value, with $b_v=121.73$ and $b_s=197.98$ MeV
respectively. The ratio $r_{S/V}=1.63$ is in good agreement with the
value derived from the previous calculations with the
non-relativistic Skyrme energy functionals~\cite{Satula06}.
Furthermore, it is shown in Fig.~\ref{fig4} that there is a strong
linear correlation (by the relation $b_s=3.28 b_v-218$ MeV) between
the volume $b_v$ and surface $b_s$ coefficients  in the symmetry
energy coefficient of finite nucleus, which is consistent with the
results in Ref.~\cite{Dieperink07}.

 \begin{figure}[tbp]
 \includegraphics[width=8cm]{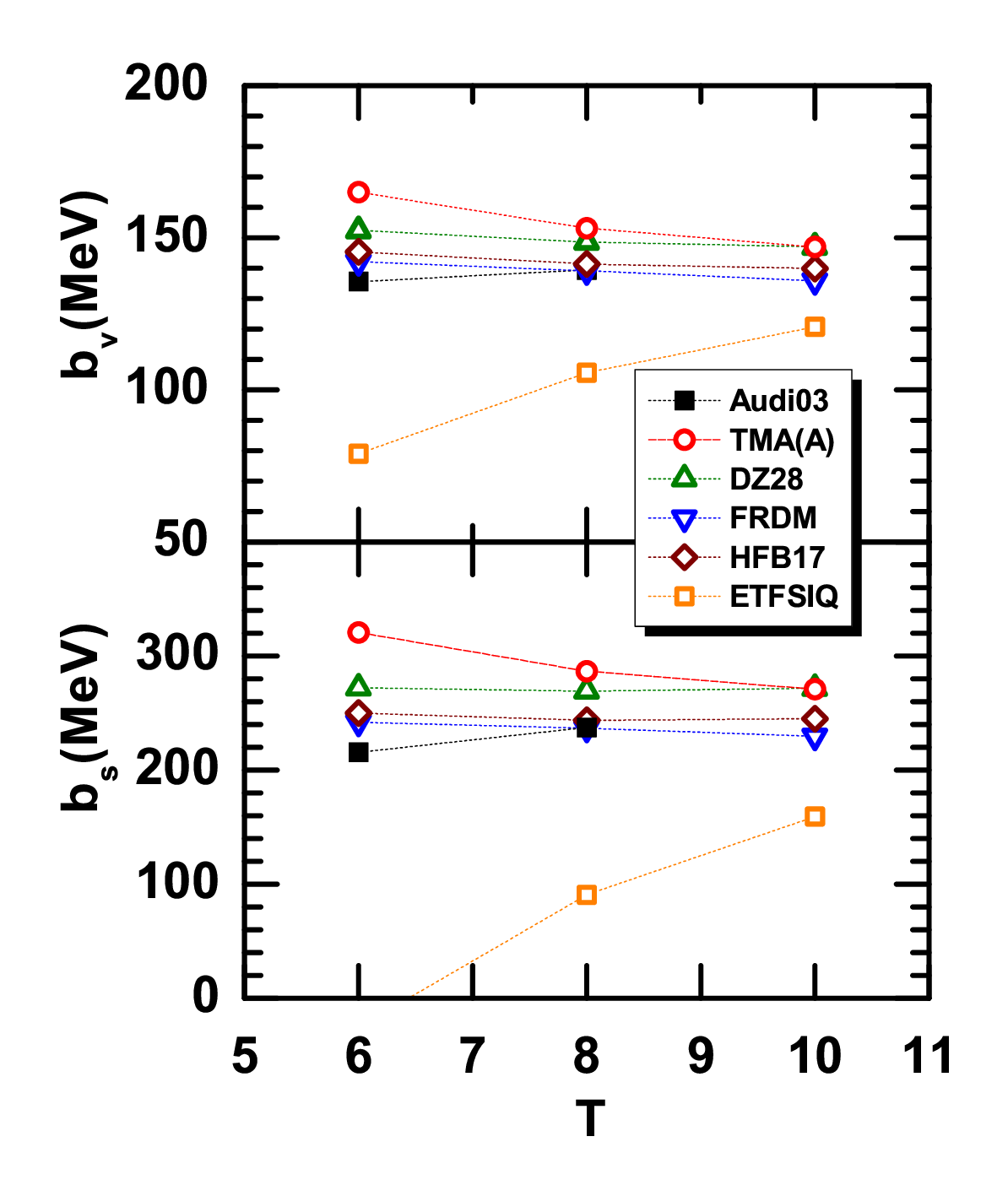}
 \caption{The optimal values of volume $b_v$ and surface $b_s$ coefficient,
  fitted to the relation (\ref{secp}) of nuclear symmetry energy
  coefficient for finite nuclei with $T=6,8,10$ in different mass tables.}
 \label{diff-T}
 \end{figure}

Furthermore, we fit the values of $b_v$ and $b_s$ to the symmetry
energy coefficients using the binding energies of nuclei in
different mass tables corresponding to smaller values of isospin,
i.e., $T=6$ and $8$. Figure~\ref{diff-T} displays the obtained $b_v$
and $b_s$ as functions of the isospin value. It is shown that except
the ``ETFSIQ", the obtained $b_s$ and $b_v$ change slowly with the
isospin value and have the tendency to be convergent at $T=10$,
which is consistent with the observation in Fig.~\ref{fig2}.

\begin{figure}[]
 \includegraphics[width=8cm]{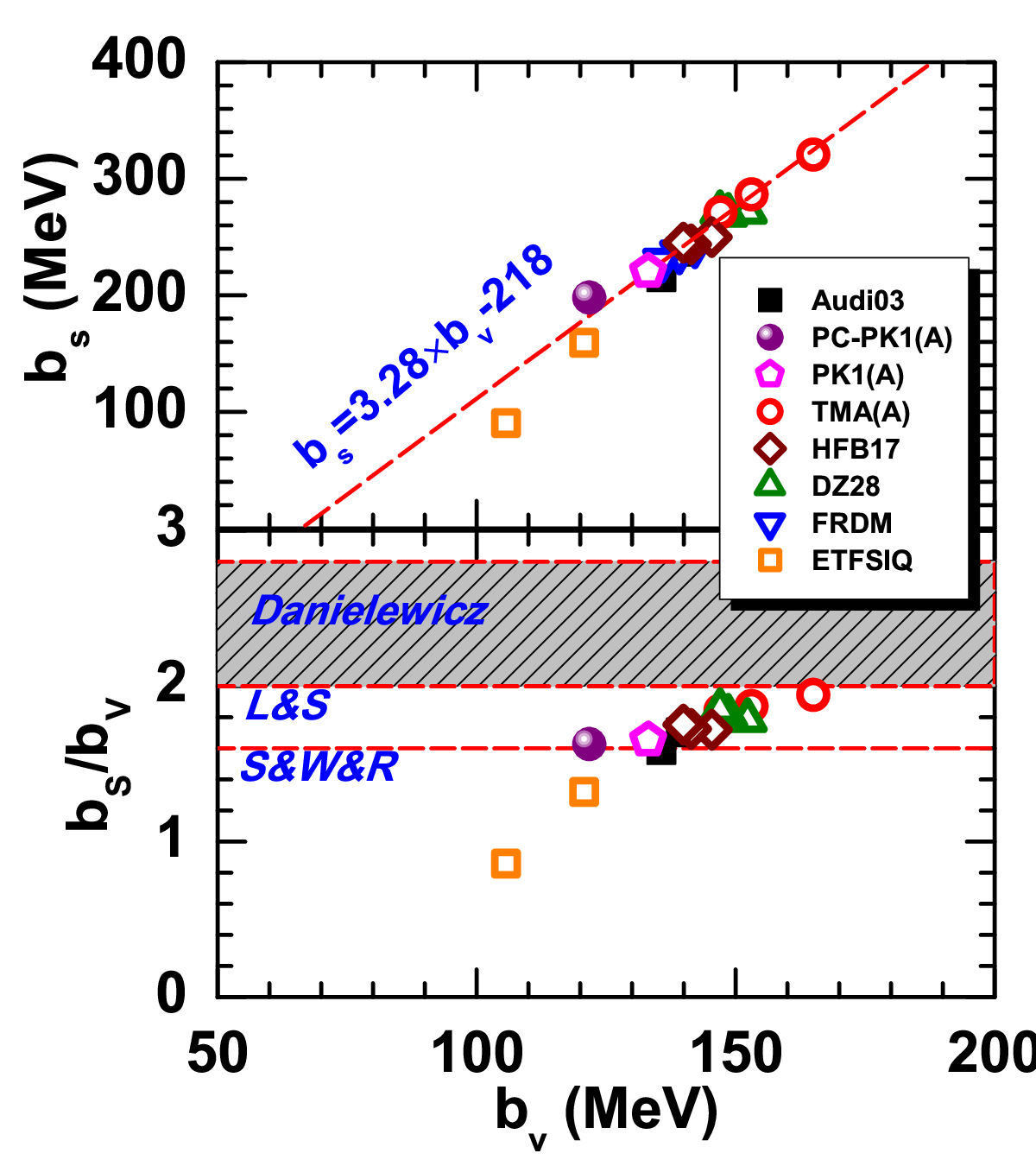}
 \caption{The surface coefficient $b_s$ and the surface-to-volume ratio $r_{S/V}(\equiv b_s/b_v)$ extracted from the
 binding energies of finite nuclei with $T=6, 8, 10$ in different mass tables
 as functions of the volume coefficient $b_v$. The results of
 axial RMF calculations with the PC-PK1 and PK1 interactions are given as well.
 The dashed lines indicates the values from the hydrodynamical model calculation~\cite{Lipparini82} and the
 SHF model calculation~\cite{Satula06}. The shade area indicates the estimation in Ref.~\cite{Danielewicz03}.}
 \label{fig6}
 \end{figure}

In Fig.~\ref{fig6}, we plot the surface coefficient $b_s$ and the
surface-to-volume ratio $r_{S/V}$ extracted from the binding
energies of finite nuclei with $T=6, 8, 10$ as functions of the
volume coefficient $b_v$.  It is shown that all the $b_s$ and $b_v$
are linearly correlated. The ratios $r_{S/V}$ of all cases, except
``ETFSIQ", are in between 1.6 and 2.0, which is consistent with the
values from the hydrodynamical model calculation~\cite{Lipparini82},
the Skryme-Hartree-Fock (SHF) model calculation in
Ref.~\cite{Satula06} as well as the result $r_{S/V}\simeq1.7$
derived from the shift of neutron-proton chemical potentials for
nuclei beyond the $\beta$-stability line with
$A\ge50$~\cite{Kolomietz10}.

\section{Summary}
 \label{SecIV}

The symmetry energy coefficient in finite nuclei have been studied
systematically in the RMF approach with the point-coupling effective
interaction PC-PK1 and compared with the values calculated using the
several available mass tables. Due to the contamination of shell
effect, the resultant nuclear symmetry energy coefficients have
large fluctuations around the nuclei with double magic numbers. This
shell effect in symmetry energy coefficient is less important for
the nuclei with larger isospin value. After subtracting the shell
effect with the Strutinsky method, the obtained nuclear symmetry
energy coefficients have been shown to decrease smoothly with the
mass number $A$. Moreover, the symmetry energy coefficients have
been fitted to the relation $\dfrac{4a_{\rm
sym}}{A}=\dfrac{b_v}{A}-\dfrac{b_s}{A^{4/3}}$. The resultant volume
$b_v$ and surface $b_s$ coefficients from axially deformed
calculations are $121.73$ and $197.98$ MeV respectively. The ratio
$r_{S/V}=1.63$ is in good agreement with the value derived from the
previous calculations with the non-relativistic Skyrme energy
functionals in Ref.~\cite{Satula06}. The coefficients $b_v$ and
$b_s$ corresponding to several available mass tables have also been
extracted. It has been shown that there is a strong linear
correlation between the volume $b_v$ and surface $b_s$ coefficients,
which is consistent with the results of previous
study~\cite{Dieperink07}. Moreover, the ratios $r_{S/V}$ have been
shown in between $1.6-2.0$ for all the cases.

 \section*{Acknowledgments}
We would like to thank S. F. Ban, J. Meng, and W. Satula for
fruitful discussions and thank Z. P. Li, P. W. Zhao and Q. B. Chen
for critical reading of this manuscript. J.M.Y. acknowledges a
postdoctoral fellowship from the F.R.S.-FNRS (Belgium). This work is
partly supported by the National Natural Science Foundation of China
under Grant No. 11105111 and No. 10947013, the Fundamental Research
Funds for the Central Universities (XDJK2010B007) and the Southwest
University Initial Research Foundation Grant to Doctor under Grant
No. SWU109011.

\end{document}